\documentclass[sigplan,nonacm]{acmart}

\usepackage{booktabs}
\usepackage{array}
\usepackage{xspace}
\usepackage{graphicx}
\usepackage{placeins}
\usepackage{stfloats}
\usepackage{amsmath}
\usepackage{xcolor}
\usepackage{fontawesome5}
\usepackage{tikz}
\usetikzlibrary{arrows.meta,positioning,shapes.geometric}

\newcommand{\dynamo}{\href{https://github.com/ai-dynamo/dynamo}{Dynamo}\xspace}
\newcommand{\snapshotrepo}{\href{https://github.com/ai-dynamo/snapshot}{Snapshot}\xspace}
\newcommand{\dynamocluster}{\href{https://github.com/ai-dynamo/dynamo}{Dynamo} cluster\xspace}

\begin{document}

\raggedbottom

\title{Fast Recovery for LLM Serving via Decoupled Device Memory Lifetime in Dynamo}

\author{Schwinn Saereesitthipitak}
\affiliation{\institution{NVIDIA}
  \country{}
}
\email{schwinns@nvidia.com}

\author{Mohammed Abdulwahhab}
\affiliation{\institution{NVIDIA}
  \country{}
}
\email{mabdulwahhab@nvidia.com}

\author{Hannah Zhang}
\affiliation{\institution{NVIDIA}
  \country{}
}
\email{hannahz@nvidia.com}

\author{Dan Feigin}
\affiliation{\institution{NVIDIA}
  \country{}
}
\email{dfeigin@nvidia.com}

\author{Neelay Shah}
\affiliation{\institution{NVIDIA}
  \country{}
}
\email{neelays@nvidia.com}

\author{Maksim Khadkevich}
\affiliation{\institution{NVIDIA}
  \country{}
}
\email{mkhadkevich@nvidia.com}

\author{Itay Neeman}
\affiliation{\institution{NVIDIA}
  \country{}
}
\email{ineeman@nvidia.com}

\author{Vikram Sharma Mailthody}
\affiliation{\institution{NVIDIA}
  \country{}
}
\email{vmailthody@nvidia.com}

\author{Wen-mei W. Hwu}
\affiliation{\institution{NVIDIA Research}
  \country{}
}
\email{whwu@nvidia.com}

\renewcommand{\shortauthors}{}

\begin{abstract}

Large language model (LLM) inference replicas run across tightly coupled GPUs and serve
traffic continuously for weeks. Hardware and software failures are therefore inevitable,
and one worker failure can disrupt an entire replica. Recovery requires reinitializing the
engine, taking minutes even when weights and compilation artifacts are cached. Production
deployments overprovision serving capacity to mask this window. We argue that the dominant
cost is loss of ready serving capacity, not request progress, so recovery should preserve
initialized engine state rather than reconstruct it.

We present fast recovery for \dynamo{} based on this principle. Snapshots capture an
initialized engine once and restore it instead of reinitializing it. Analysis of 18 weeks
of failures from the \dynamocluster{} shows that most failures are
\emph{device-preserving}: the engine process fails while the GPU and its resident
allocations remain intact. Our key insight is that independent engine processes can reuse
the same GPU-resident state while keeping mutable execution state private. The GPU Memory
Service (GMS) decouples device-memory ownership from engine processes, enabling engines to
share and reattach surviving allocations without copying them. GMS preserves model weights
and shares them read-only between replacement and Shadow Engines, avoiding weight reloads.
A second initialized runtime on the same GPUs reduces recovery to promotion. Across four
models on vLLM and SGLang, these mechanisms recover a failed replica in under 7 seconds,
13--29 times faster than a warm restart, using a fixed 4--8 GiB of device memory per GPU
independent of model size. Replaying the production trace, we estimate they would reclaim
79\% of GPU-hours lost to recovery.

\end{abstract}
 
\maketitle

\section{Introduction}
\label{sec:intro}

Modern LLM inference has evolved into a large-scale distributed systems workload, with
a single serving endpoint spanning dozens to hundreds of tightly coupled GPUs.
Each endpoint comprises several homogeneous replicas of the same model across a large GPU cluster, which may also host many endpoints.
A replica is an instance of an inference engine that typically comprises several to dozens of GPUs,
optimized for maximum throughput subject to latency and interactivity objectives~\cite{distserve-osdi24,sarathi-osdi24,dynamo}. Replicas often sustain
hundreds of thousands of tokens per second and exceed one million tokens
per second on rack-scale systems~\cite{deepseek-inference-overview,nvidia-gptoss-nvl72}.
These engines are long-running, continuously serving traffic for weeks.
Consequently, even infrequent component failures (software or hardware) become inevitable over an engine’s
lifetime. Moreover, because GPUs within an engine are tightly coupled, a failure of a
single worker can disrupt the entire replica.

\begin{figure}[t]
  \centering
  \includegraphics[width=\linewidth]{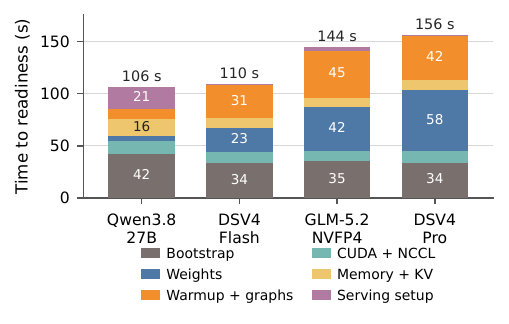}
  \caption{Restart time for vLLM 0.27.1 on 8xB200 GPUs (TP/TEP8). Compilation artifacts are pre-cached and weights pre-pulled into page cache.}
  \label{fig:startup}
\end{figure}

Recovering a failed replica requires restarting the inference engine, a process that can take several minutes
even when all caches are warm (Figure~\ref{fig:startup}). During this recovery window, in-flight requests are re-queued,
and new traffic is redirected to the remaining replicas. The resulting capacity loss increases admission delays
and degrades interactivity, ultimately leading to service-level objective (SLO) violations. Under sustained
load, the remaining replicas may be unable to absorb the displaced traffic, causing requests to be dropped
entirely. Meanwhile, the recovering replica continues to occupy provisioned GPUs without producing tokens, wasting valuable GPU capacity (Figure~\ref{fig:gpu-hours}).

Fault tolerance has been extensively studied for large-scale training~\cite{megascale-nsdi24}.
Existing approaches bound lost computation through checkpointing~\cite{checkfreq-fast21,gemini-sosp23}, maintain
execution through redundant resources~\cite{bamboo-nsdi23,oobleck-sosp23}, or elastically remap computation onto
surviving resources~\cite{varuna-eurosys22,recycle-sosp24}. These techniques fundamentally aim to preserve
training progress: model weights and optimizer states evolve with every iteration and encode the computation
performed thus far. Prior work has applied a similar principle to inference by preserving request progress,
either by replicating the KV cache~\cite{dejavu-icml24,failsafe-arxiv25} or by reconfiguring parallelism to
continue execution despite failed workers~\cite{tarragon-arxiv26,revivemoe-arxiv26}.

For inference, however, the cost of failure is fundamentally different: the dominant loss is not request
progress, but \textbf{ready serving capacity}. Before a replica can serve requests, every rank must load its
model shard and construct the host and device execution environment, including collective state, specialized
kernels, tuning decisions, and CUDA graphs (Section~\ref{sec:anatomy}). This initialization is independent of
request traffic, yet conventional recovery reconstructs much of this expensive state from scratch
(Figure~\ref{fig:startup}). In contrast, the transient state associated with an interrupted request—its
scheduler and KV state—can be reconstructed by replaying the request’s prefill, typically within seconds. Thus,
\textbf{assuming failures can be detected promptly, the primary recovery objective for inference should be restoring
ready serving capacity, rather than preserving individual request progress.}

Production inference deployments routinely reserve spare GPU capacity to mask the multi-minute recovery time of
failed replicas. Rather than placing restart latency on the critical path, operators overprovision the serving
fleet so that healthy replicas can absorb displaced traffic while a replacement becomes ready. This is a
pragmatic response to expensive engine initialization: large-model startup can take several minutes even when
model artifacts are locally
available in nodes' SSDs, as engines must load weights, initialize distributed state, compile and
tune kernels, and capture CUDA graphs~\cite{nvidia-cuda-graphs}. The cost, however, is substantial: overprovisioned GPUs are
paid for continuously, despite being needed only during transient failures.

This strategy is particularly attractive because it provides a uniform recovery mechanism for both process
failures and hardware loss. However, our production failure analysis reveals an important asymmetry: most
failures observed over 18 weeks are \emph{device-preserving}. Although the inference
engine process fails, the underlying GPU remains healthy, and unrelated resident allocations remain intact and
trustworthy across the failure boundary (Section~\ref{sec:failures}).
This observation exposes a key
opportunity: \emph{if the lifetime of a replica’s committed resident state is decoupled from that of the engine
process, this state can survive engine failures and be reused during recovery.}

Together, these observations motivate a fast-recovery architecture for \textbf{\dynamo{}}, built from three complementary mechanisms:

\begin{itemize}
\item \emph{\href{https://github.com/ai-dynamo/snapshot}{Snapshots}} (Section~\ref{sec:design-restore}) eliminate repeated framework initialization.
An initialized engine is checkpointed once, before entering service, and subsequently restored after failure.
Recovery therefore replaces a complex and compute-intensive initialization path with predictable state
restoration and data movement, without requiring shadow resources.

\item \emph{GPU Memory Service (GMS)}
(Section~\ref{sec:design-substrate}) eliminates model reloading after device-preserving failures.
This per-GPU service owns committed model allocations independently of engine processes.
Using CUDA virtual memory management, engines import allocation handles and map the same physical
model state read-only into their own address spaces, without copying model bytes or adding
indirection to execution. A replacement engine can therefore reattach surviving model state.

\item \emph{Shadow Engines} (Section~\ref{sec:design-reattach}) remove runtime initialization
from the promotion path. GMS lets a shadow share the primary's model weights on the same GPUs
while retaining an independent, initialized host and device runtime. This avoids a second model
copy and reduces recovery to promotion and installation of fresh KV backing.

\end{itemize}

For device-preserving failures, \dynamo{}'s GMS and Shadow Engines directly reuse state that survives on the original
GPUs, avoiding reconstruction of model and engine state.
Device-invalidating failures instead fall back to a
durable Snapshot, which restores the engine on a healthy device while GPU-resident state is reconstructed. When composed, the three mechanisms form a tiered recovery path: a resident shadow
provides near-immediate recovery for the common case, and Snapshots replenish consumed shadows outside the
serving critical path, while providing a recovery fallback for broader failures. Table~\ref{tab:composition-spectrum}
describes this composition and the resulting design space, which trades storage and memory headroom for
progressively lower recovery latency.

Our evaluation shows that a Shadow Engine in \dynamo{} recovers a replica experiencing a device-preserving failure in under 7 seconds, 13--29$\times$ faster than a warm restart, at the expense of 4--8 GiB of per-GPU memory occupation overhead.
Under a device-invalidating failure, Snapshots (when used with GMS) can recover an engine 2.1--5.4$\times$ faster than warm restart.
A counterfactual analysis of 18 weeks of production failures shows that reducing recovery to 7 seconds with Shadow Engines would have reclaimed 79\% of all GPU-hours lost to recovery, counting savings only for failures we can confirm as device-preserving.

Overall, this work makes the following key contributions:
\begin{itemize}
\item \textbf{Production failure characterization.} We characterize inference-engine failures and their recovery costs using large-scale production data, including the prevalence and properties of device-preserving and device-invalidating failures (Section~\ref{sec:failures}).

\item \textbf{A composable recovery architecture.} We design a unified recovery architecture that combines serialization and residency to preserve initialized engine state across failures (Section~\ref{sec:design}).

\item \textbf{End-to-end implementation and evaluation.} We implement all proposed mechanisms in a production-scale inference stack and evaluate them under controlled failure injection, quantifying recovery latency, shadow memory overhead, TTFT and ITL during replica failure, serving behavior during replenishment, and shadow replenishment time (Section~\ref{sec:eval}).
\end{itemize}

\begingroup
\urlstyle{same}
\small
\noindent The open-source implementation is available at
\url{https://github.com/ai-dynamo/dynamo} and
\url{https://github.com/ai-dynamo/snapshot}.
\endgroup

\section{Background and Motivation}
\label{sec:motivation}

\subsection{Inference Engine Startup Sequence}
\label{sec:anatomy}

An inference engine must complete a sequence of initialization steps on every worker rank before it can serve
requests. Each worker first creates its device execution context (e.g., a CUDA context), followed by collective
communication state, including host- and device-side handles, channels, and buffers. The engine then constructs
the model and loads each rank’s weight shard into device memory. Representative prefill and decode passes
initialize or specialize kernels, select tuned implementations, and capture CUDA graphs for supported shape and
batch-size buckets. Finally, the engine profiles its peak device-memory footprint and reserves the remaining
capacity for the KV-cache pool. On restart, some artifacts may already be cached—for example, model weights in
the host page cache and compiled kernel artifacts—but much of the process- and device-local state must still be
reconstructed before the engine can serve requests.

Consider the hypothetical scenario of two engine instances initialized with the same model configuration and
topology. The rank-local weight tensors have identical shapes and contents after initialization. We refer to
these tensors as \emph{model state}. Initialization also creates \emph{runtime state}: the host and device state
required to execute the inference loop. Device runtime state includes the CUDA context, CUDA graphs,
communicators, and execution buffers. Because initialization is deterministic for a fixed configuration,
topology, and hardware~\cite{foundry-arxiv26}, engine instances have the same runtime-state structure, including
CUDA graphs and device virtual-address layouts~\cite{nvidia-cuda-graphs,nvidia-cuda-vmm}. The underlying
physical allocations and deployment-specific host state, such as network endpoints and communication handles,
may differ. Importantly, host and device runtime state can be coupled. For example, NCCL operations captured in
CUDA graphs depend on host-side communication state~\cite{nccl}.

Finally, \emph{request state} is traffic-dependent and includes host-side scheduler metadata and KV mappings~\cite{vllm-sosp23}, and device-side KV-cache contents.
Unlike model and runtime state, request state depends on the traffic processed by
an engine.

\subsection{Failure Model}
\label{sec:failures}

A production inference deployment consists of many replicas, each comprising multiple processes spanning several
to dozens of GPUs. Because ranks execute collectively, a failure in one rank can stall the entire replica,
requiring all ranks to restart.
We classify failures by what recovery must rebuild. A failure is \emph{device-preserving} if the GPUs need no
reset and the resident state of other processes on them remains intact and trustworthy, so that a replacement
engine can reuse it. A failure is \emph{device-invalidating} if a GPU must be reset or replaced, or a
memory-integrity error makes its contents suspect, so that resident state must be reconstructed.

Most engine failures raise no XID: hangs, out-of-memory errors, or exceptions can interrupt serving while GPUs
remain schedulable. However, neither schedulability nor the absence of XIDs guarantees that resident state is
safe to reuse. When reuse is uncertain, recovery falls back to checkpoint restoration and state reconstruction on a
healthy device, which may be the original GPU after remediation. Application faults (XIDs~13, 31, and 43) generally permit
application restart~\cite{nvidia-xid-catalog}, whereas memory-integrity errors (XIDs~48, 94, and 95) and reset-class
faults (XIDs~32, 79, 109, and 119) invalidate resident-state reuse
under our policy~\cite{nvidia-gpu-error-containment,nvidia-gpu-recovery-actions}. Appendix Table~\ref{tab:relevant-xids}
summarizes these signals and recovery actions.

\paragraph{Production failure distribution.}

\begin{figure}[t]
  \centering
  \includegraphics[width=\linewidth]{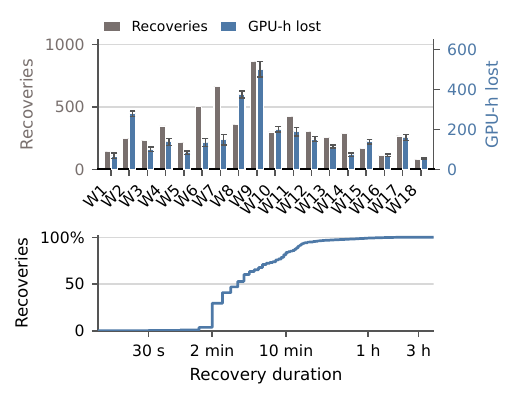}
  \caption{Recoveries in the \dynamocluster{} over 18 weeks: recoveries per week and the
  GPU-hours lost to them (top), and the distribution of device-preserving
  recovery duration, from container restart to ready (bottom). Whiskers show
  timing uncertainty (Appendix~\ref{app:trace-timing}).}
  \Description{Two stacked panels. The top panel is a grouped bar chart of
  weekly serving-worker recoveries on the left axis and idle GPU-hours with
  small error bars on the right axis over 18 anonymized weeks. The bottom
  panel is a cumulative distribution of recovery duration on a logarithmic
  axis, rising steeply between two and ten minutes.}
  \label{fig:gpu-hours}
\end{figure}

\begin{table}[t]
  \small
  \centering
  \caption{Failures in the \dynamocluster{} over 18 weeks. Engine failures count replica
  restarts, excluding crash loops. The other rows count episodes.}
  \label{tab:failure-incidents}
  \begin{tabular*}{\columnwidth}{@{\extracolsep{\fill}}p{0.45\columnwidth}p{0.18\columnwidth}p{0.28\columnwidth}@{}}
    \toprule
    \raggedright\textbf{Failure / signal} & \centering\textbf{Count} &
    \raggedright\textbf{Recovery action} \tabularnewline
    \midrule
    \raggedright Engine failure, device-preserving
      & \centering 4{,}443
      & \raggedright Restart replica. \tabularnewline
    \addlinespace
    \raggedright Engine failure, ambiguous
      & \centering 1{,}319
      & \raggedright Restart replica. \tabularnewline
    \addlinespace
    \raggedright GPU hang (XIDs~32/109)
      & \centering 2
      & \raggedright Reset affected GPUs. \tabularnewline
    \addlinespace
    \raggedright Contained memory error (XID~94)
      & \centering 1
      & \raggedright Reset affected GPUs. \tabularnewline
    \bottomrule
  \end{tabular*}
\end{table}

Over 18 weeks, we collected production failure data from one of the \dynamo{} clusters. The measured cluster comprises
783 NVIDIA B200 and 135 NVIDIA RTX PRO 6000 Blackwell GPUs. The service hosts a mix of LLM and diffusion
models across approximately 150 concurrent inference-engine replicas. Each replica typically spans 8 GPUs
(a B200 node) or fewer depending on the served model. We observed 5,762 in-place restarts of serving
replicas, an average of 46 per day, with a median recovery time of 3.5 minutes and 2,967 GPU-hours lost to
recovery alone, excluding capacity reserved through overprovisioning.

\textit{Most observed failures were device-preserving.} To be conservative, we count a restart as
device-preserving only if the replica returned to service on the same GPUs, the node lost no schedulable GPU
meanwhile, and no reset-class or memory-integrity XID was reported on that host in the same week.
These retrospective criteria estimate the opportunity for state reuse.
They do not establish its safety at recovery time.
4,443 restarts meet these conditions, accounting for 2,375 GPU-hours with a median recovery of 4 minutes
(Figure~\ref{fig:gpu-hours}). The remaining 1,319 are ambiguous and excluded from our analysis
(Table~\ref{tab:failure-incidents}). Device-invalidating failures, such as GPU hangs that required a device
reset (XIDs~32/109) and contained memory errors (XID~94), were rare. Thus, production recovery reconstructs an entire
inference engine even though the underlying GPUs, and potentially reusable resident state, remain available.

\section{Design}
\label{sec:design}

\begin{table*}[!b]
  \caption{Comparison of recovery paths for device-preserving failures.}
  \label{tab:composition-spectrum}
  \centering
  \scriptsize
  \setlength{\tabcolsep}{3pt}
  \renewcommand{\arraystretch}{1.20}
  \begin{tabular}{@{}
    >{\raggedright\arraybackslash}p{0.16\textwidth}
    >{\raggedright\arraybackslash}p{0.22\textwidth}
    >{\raggedright\arraybackslash}p{0.39\textwidth}
    >{\raggedright\arraybackslash}p{0.17\textwidth}@{}}
    \toprule
    \textbf{Configuration} & \textbf{State to recover} &
      \textbf{Recovery action} & \textbf{Extra device memory} \\
    \midrule
    Warm restart
      & Model state, host \& device runtime, request state
      & Reinitialize runtime and load model state
      & None \\
    Snapshot
      & Model state, host \& device runtime, request state
      & Restore Snapshot
      & None \\
    GMS
      & Host \& device runtime, request state
      & Reinitialize runtime and attach model state read-only
      & None \\
    Snapshot + GMS
      & Host \& device runtime, request state
      & Restore runtime and attach model state read-only
      & None \\
    GMS + Shadow Engine
      & Request state
      & Promote Shadow Engine and install KV backing, replenish shadow
      & Shadow device runtime \\
    All three
      & Request state
      & Promote Shadow Engine and install KV backing, restore shadow from Snapshot
      & Shadow device runtime \\
    \bottomrule
  \end{tabular}
  \Description{Six configurations are compared by the state or resources that
  must be recovered after a device-preserving failure, the corresponding
  recovery action, and additional device-memory occupancy.}
\end{table*}

\definecolor{sThreeModel}{HTML}{B9D9EE}
\definecolor{sThreeModelEdge}{HTML}{286B91}
\definecolor{sThreeHost}{HTML}{F6C58F}
\definecolor{sThreeHostEdge}{HTML}{A65712}
\definecolor{sThreeDevice}{HTML}{F3DB91}
\definecolor{sThreeDeviceEdge}{HTML}{8B6B12}
\definecolor{sThreeRequest}{HTML}{E6E6E6}
\definecolor{sThreeFailure}{HTML}{B73535}
\tikzset{
  s3panel/.style={draw=black!45, line width=0.45pt, rounded corners=1pt},
  s3box/.style={draw=black!75, line width=0.55pt, rounded corners=1pt,
    align=center, minimum height=0.56cm, inner sep=2.5pt, font=\scriptsize},
  s3model/.style={s3box, fill=sThreeModel, draw=sThreeModelEdge},
  s3host/.style={s3box, fill=sThreeHost, draw=sThreeHostEdge},
  s3device/.style={s3box, fill=sThreeDevice, draw=sThreeDeviceEdge},
  s3devicegroup/.style={draw=sThreeDeviceEdge, fill=sThreeDevice!28,
    line width=0.6pt, rounded corners=1pt},
  s3absent/.style={draw=sThreeFailure, densely dotted, line cap=round},
  s3devicegroupmissing/.style={s3devicegroup, s3absent},
  s3devicegroupdead/.style={s3devicegroup, s3absent},
  s3runtime/.style={s3host},
  s3address/.style={s3device},
  s3request/.style={s3box, fill=sThreeRequest, draw=black!55},
  s3checkpoint/.style={draw=black!65, line width=0.55pt, double,
    rounded corners=1pt},
  s3modelmissing/.style={s3model, s3absent},
  s3hostmissing/.style={s3host, s3absent},
  s3devicemissing/.style={s3device, s3absent},
  s3requestmissing/.style={s3request, s3absent},
  s3modeldead/.style={s3model, s3absent},
  s3hostdead/.style={s3host, s3absent},
  s3devicedead/.style={s3device, s3absent},
  s3requestdead/.style={s3request, s3absent},
  s3mapping/.style={-{Latex[length=1.35mm,width=0.95mm]},
    line width=0.75pt, draw=black!75, densely dotted, line cap=round},
  s3flow/.style={-{Latex[length=1.55mm]}, line width=0.65pt,
    draw=black!75},
  s3step/.style={circle, fill=black!80, text=white, inner sep=0pt,
    minimum size=0.34cm, font=\tiny\bfseries}
}
\newcommand{\sThreeStep}[1]{\tikz[baseline=(s3inline.base)]{\node[circle, draw=black!70, line width=0.4pt, fill=white,
      inner sep=0pt, minimum size=1.25em, font=\scriptsize\bfseries]
      (s3inline) {#1};}}
\newcommand{\sThreeLock}{\raisebox{-0.14em}{\fontsize{8pt}{8pt}\selectfont\faLock}}

Dynamo's recovery architecture accelerates inference recovery by preserving initialized engine state across failures rather than reconstructing it. Section~\ref{sec:anatomy} separates an initialized engine into model, runtime, and request state. The architecture focuses on model and runtime state, which are independent of request traffic and can therefore be prepared before serving and reused after failure. Request state is orthogonal to these mechanisms. Our implementation replays interrupted requests, as discussed in Section~\ref{sec:impl}.

The architecture provides three mechanisms that progressively remove work from the
recovery path. \emph{Snapshots} serialize an initialized engine into
a reusable artifact, replacing framework initialization with state restoration
(Figure~\ref{fig:snapshot-protocol}). \emph{GMS} decouples model
allocations from the engine process, enabling committed model state to remain
resident across device-preserving failures and be reattached by a replacement
engine. Finally, \emph{Shadow Engines} keep runtime state initialized alongside
the primary engine while sharing the same GMS-owned model state. Recovery then
requires only promoting the shadow, provisioning fresh KV pages, and replaying
interrupted requests.

These mechanisms compose into a tiered recovery path. For the common
device-preserving failure, a shadow retains both runtime state and access to
resident model state, providing the fastest recovery. If no shadow is
available, a Snapshot can restore the runtime while GMS preserves
the model state. For device-invalidating failures, where resident state cannot
be reused, the durable Snapshot provides the fallback and reconstructs
device state on a healthy GPU, while the GMS is populated in parallel to accelerate recovery time. The resulting configurations trade storage
and memory capacity for progressively less work on the recovery critical
path. Table~\ref{tab:composition-spectrum} summarizes the
resulting device-preserving recovery paths and memory tradeoffs.

\subsection{Snapshots}
\label{sec:design-restore}

\begin{figure}[t]
  \centering
  \begin{tikzpicture}[
    pcbox/.style={minimum height=0.34cm, inner sep=1.2pt,
      font=\fontsize{5.2}{5.6}\selectfont},
    pctitle/.style={font=\scriptsize\bfseries},
    pcflowlabel/.style={font=\fontsize{5.2}{5.6}\selectfont,
      fill=white, inner sep=0.7pt},
    pchostgroup/.style={draw=sThreeHostEdge, fill=sThreeHost!28,
      line width=0.6pt, rounded corners=1pt},
    pcstorage/.style={cylinder, shape border rotate=90, aspect=0.28,
      draw=black!65, fill=white, line width=0.55pt,
      minimum width=2.55cm, minimum height=0.95cm, align=center,
      font=\scriptsize\bfseries}
  ]
\begin{scope}[shift={(-2.05,1.30)}]
      \draw[s3panel] (-1.57,-1.18) rectangle (1.57,1.30);
      \node[pctitle] at (0,1.08) {Warm engine};
      \node[s3step] at (-1.36,1.08) {1};
      \node[s3host, pcbox, minimum width=2.68cm]
        at (0,0.67) {Host runtime};
      \draw[s3devicegroup] (-1.36,-0.92) rectangle (1.36,0.38);
      \node[font=\fontsize{5.4}{5.8}\selectfont\bfseries]
        at (0,0.20) {Device state};
      \node[s3device, pcbox, minimum width=1.18cm]
        at (-0.65,-0.20) {Runtime};
      \node[s3model, pcbox, minimum width=1.18cm]
        at (0.65,-0.20) {Model pages};
      \node[s3device, pcbox, minimum width=2.48cm]
        at (0,-0.68) {KV pages};
    \end{scope}

    \begin{scope}[shift={(2.05,1.30)}]
      \draw[s3panel] (-1.57,-1.18) rectangle (1.57,1.30);
      \node[pctitle] at (0,1.08) {Prepared engine};
      \node[s3step] at (-1.36,1.08) {2};
      \node[s3host, pcbox, minimum width=2.68cm]
        at (0,0.67) {Host runtime};
      \draw[s3devicegroup] (-1.36,-0.92) rectangle (1.36,0.38);
      \node[font=\fontsize{5.4}{5.8}\selectfont\bfseries]
        at (0,0.20) {Device state};
      \node[s3device, pcbox, minimum width=1.18cm]
        at (-0.65,-0.20) {Runtime};
      \node[s3model, pcbox, minimum width=1.18cm]
        at (0.65,-0.20) {Model pages};
      \node[s3devicemissing, pcbox, minimum width=2.48cm]
        at (0,-0.68) {KV pages};
    \end{scope}

\draw[s3flow] (-0.48,1.58) -- node[pcflowlabel, above] {Prune} (0.48,1.58);
    \draw[s3flow] (0.48,1.02) -- node[pcflowlabel, below] {Remap} (-0.48,1.02);

    \draw[s3panel] (0.48,-1.68) rectangle (3.62,-0.40);
    \node[pctitle] at (2.05,-0.62) {Host process};
    \node[s3step] at (0.70,-0.62) {3};
    \draw[pchostgroup] (0.65,-1.58) rectangle (3.45,-0.82);
    \node[font=\fontsize{5.2}{5.6}\selectfont\bfseries]
      at (2.05,-0.97) {Host runtime};
    \node[s3device, pcbox, minimum width=2.45cm]
      at (2.05,-1.34) {Device state};

    \draw[s3flow] (2.32,0.12) -- node[pcflowlabel, right] {Capture (Device)} (2.32,-0.40);
    \draw[s3flow] (1.78,-0.40) -- node[pcflowlabel, left] {Restore (Device)} (1.78,0.12);

    \node[pcstorage] (storage) at (-2.05,-1.04) {Snapshot};
    \node[s3step] at (-3.55,-0.78) {4};
    \draw[s3flow] (0.48,-0.90) -- node[pcflowlabel, above] {Capture (Host)} (-0.77,-0.90);
    \draw[s3flow] (-0.77,-1.18) -- node[pcflowlabel, below] {Restore (Host)} (0.48,-1.18);

    \draw[draw=black!65, line width=0.55pt]
      (-1.56,-2.02) rectangle (-1.35,-1.85);
    \node[font=\tiny, anchor=west] at (-1.25,-1.935) {present};
    \draw[s3absent, line width=0.55pt]
      (0.12,-2.02) rectangle (0.33,-1.85);
    \node[font=\tiny, anchor=west] at (0.43,-1.935) {absent / lost};
  \end{tikzpicture}
  \caption{Snapshot capture and restore. Capture releases empty KV pages, stages remaining device state in the host runtime, and checkpoints the process. Restore reloads the process and device state and remaps fresh KV pages.
}
  \Description{A warm engine is pruned into a prepared engine by releasing its
  empty KV pages. In the next stage, the host process contains the host runtime,
  which contains the entire staged device state. The host process is saved as a
  Snapshot. Reverse arrows load the process, restore the device state, and
  remap fresh KV pages.}
  \label{fig:snapshot-protocol}
\end{figure}
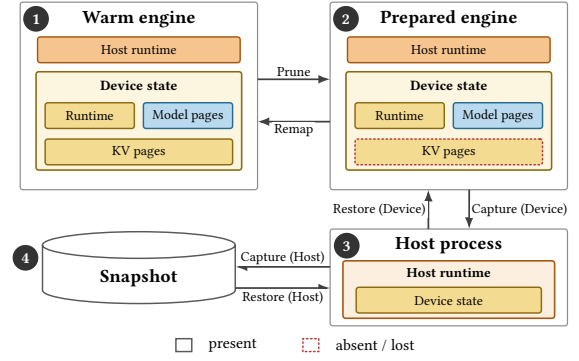

A \emph{Snapshot} is a durable snapshot of an initialized engine.
It captures the host process tree, including address spaces, threads,
registers, and file-descriptor metadata, together with CUDA state such as
contexts, execution state, virtual-address mappings, and referenced GPU
allocations~\cite{criu,cuda-checkpoint}.

Figure~\ref{fig:snapshot-protocol} shows capture and restore in \dynamo{}.
\sThreeStep{1}\ The engine fully initializes and warms up before accepting
traffic. \sThreeStep{2}\ It prunes device memory by releasing disposable
allocations and detaching externally owned ones, while retaining required
process-owned state. \sThreeStep{3}\ CUDA execution is frozen, and the remaining
device state is staged into host memory. \sThreeStep{4}\ The host process tree
and staged device state are serialized to durable storage. Recovery reverses
these steps: the process tree is restored, CUDA state is reinstated, released
or detached allocations are recreated or re-imported, and serving resumes.
Because the checkpoint is durable, it can also recover device-invalidating
failures on compatible devices and rank topologies.

Checkpoint recovery is dominated by moving the serialized state. For an
artifact of size $S$, the data transfer component of recovery latency is approximately
$S/B_{\mathrm{eff}}$, where $B_{\mathrm{eff}}$ is the effective restore
bandwidth. Thus, recovery improves either by increasing transfer bandwidth or
reducing $S$. We reduce $S$ by excluding state that need not be checkpointed. The checkpoint
is captured before request admission, so the KV cache contains no request data
even though its physical pool may occupy most remaining GPU memory. We release
these KV pages while retaining their virtual-address reservations, preserving
addresses embedded in CUDA graphs and tensor metadata. Fresh KV pages are
remapped at the same addresses after restore.

Model weights are typically the largest remaining component of $S$. GMS
(Section~\ref{sec:design-substrate}) moves these allocations outside the engine
process, allowing them to remain resident while being detached before capture.
Section~\ref{sec:design-gms-pc} describes this composition, which removes model
weights from the checkpoint artifact as well.

\subsection{GMS: GPU Memory Service}
\label{sec:design-substrate}

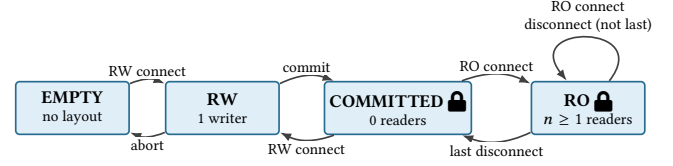
\begin{figure}[t]
  \centering
  \resizebox{\columnwidth}{!}{\begin{tikzpicture}[
    gmsstate/.style={draw=sThreeModelEdge, fill=sThreeModel!45,
      line width=0.65pt, rounded corners=1.5pt, align=center,
      minimum width=1.55cm, minimum height=0.72cm, inner sep=2pt,
      font=\scriptsize\bfseries},
    gmsedge/.style={s3flow},
    gmslabel/.style={font=\tiny, fill=white, inner sep=1pt, align=center}
  ]
    \node[gmsstate] (empty) at (0,0) {EMPTY\\[-1pt]
      \normalfont\tiny no layout};
    \node[gmsstate] (rw) at (2.05,0) {RW\\[-1pt]
      \normalfont\tiny 1 writer};
    \node[gmsstate, minimum width=1.90cm] (committed) at (4.45,0)
      {COMMITTED \sThreeLock\\[-1pt]\normalfont\tiny 0 readers};
    \node[gmsstate] (ro) at (7.05,0) {RO \sThreeLock\\[-1pt]
      \normalfont\tiny $n \geq 1$ readers};

    \draw[gmsedge] (empty) to[bend left=23]
      node[gmslabel, above] {RW connect} (rw);
    \draw[gmsedge] (rw) to[bend left=23]
      node[gmslabel, below] {abort} (empty);
    \draw[gmsedge] (rw) to[bend left=23]
      node[gmslabel, above] {commit} (committed);
    \draw[gmsedge] (committed) to[bend left=23]
      node[gmslabel, below] {RW connect} (rw);
    \draw[gmsedge] (committed) to[bend left=23]
      node[gmslabel, above] {RO connect} (ro);
    \draw[gmsedge] (ro) to[bend left=23]
      node[gmslabel, below] {last disconnect} (committed);
    \draw[gmsedge] (ro) to[out=52,in=128,looseness=4.4]
      node[gmslabel, above, yshift=1pt] {RO connect\\disconnect (not last)} (ro);
  \end{tikzpicture}}
  \caption{The GMS model-state FSM. \textsc{Committed} and \textsc{RO}
  refer to the same committed model state. They differ only in whether a reader
  lease is attached.}
  \Description{Four states form the implemented GMS state machine: Empty,
  exclusive read-write, committed with no readers, and shared read-only with
  one or more readers. Commit publishes model state, reader connections and
  disconnections change the reader count, abort clears an incomplete write, and
  a writer may reacquire and reuse the allocation set after all readers leave.}
  \label{fig:gms-fsm}
\end{figure}

Conventionally, model allocations are owned by the engine's CUDA context and
are destroyed with it. GMS decouples their lifetime from the engine by owning
physical allocations in a separate process on each GPU and exporting their
handles to engines over Unix-domain sockets. GMS retains the allocation
handles but never maps or accesses the pages. Thus, committed model state
survives a device-preserving engine failure as long as GMS and the GPU remain
healthy.

Figure~\ref{fig:gms-fsm} shows the ownership protocol. A cold-starting engine
acquires an exclusive read-write lease, populates the GMS allocations, and
commits them. Commit makes the model state immutable. The writer subsequently
reconnects read-only, and additional engines may attach as readers. If a writer
fails before commit, GMS discards the incomplete state. Reader failures only
release their leases, leaving committed state intact. Exclusive write access
becomes available again after the last reader disconnects.

GMS enforces read-only access using \texttt{cuMemSetAccess}. An invalid write
from an engine triggers a GPU MMU fault (XID~31) before modifying the shared
pages, isolating the fault to the offending engine while preserving the
GMS-owned model state~\cite{nvidia-cuda-vmm,nvidia-xid-catalog}. Integrity
errors or device loss invalidate the corresponding GMS state and follow the
device-invalidating recovery path~\cite{nvidia-gpu-error-containment,
nvidia-dynamic-page-offlining}.

After an engine failure, a replacement reconstructs its runtime state but
avoids reloading model weights. It acquires a read-only lease, imports the GMS
allocations, and maps them into its CUDA context's virtual address space.
During reconstruction,
host-side model initialization is replayed while device operations are
suppressed until the existing model state is attached.

\subsection{Snapshots with GMS}
\label{sec:design-gms-pc}

\begin{figure}[t]
  \centering
  \begin{tikzpicture}[
    mapbox/.style={minimum height=0.34cm, inner sep=1.1pt, font=\tiny},
    maptitle/.style={font=\scriptsize\bfseries},
    maplabel/.style={font=\tiny, fill=white, inner sep=0.8pt}
  ]
\node[maptitle] at (-2.85,1.42) {(a) Populate};
    \draw[s3panel] (-3.80,-0.05) rectangle (-1.90,1.15);
    \node[font=\tiny\bfseries] at (-2.85,0.95) {Writer};
    \draw[s3devicegroup] (-3.63,0.06) rectangle (-2.07,0.72);
    \node[font=\fontsize{5.1}{5.5}\selectfont\bfseries]
      at (-2.85,0.58) {Device state};
    \node[s3address, mapbox, minimum width=1.20cm] (map-writer-va)
      at (-2.85,0.30) {Model VA};
    \node[s3model, mapbox, minimum width=1.68cm] (map-writer-gms)
      at (-2.85,-0.70) {GMS pages};
    \draw[s3mapping] (map-writer-va.south) --
      node[maplabel, right] {RW} (map-writer-gms.north);

    \node[maptitle] at (0,1.42) {(b) Serve};
    \draw[s3panel] (-0.95,-0.05) rectangle (0.95,1.15);
    \node[font=\tiny\bfseries] at (0,0.95) {Reader};
    \draw[s3devicegroup] (-0.78,0.06) rectangle (0.78,0.72);
    \node[font=\fontsize{5.1}{5.5}\selectfont\bfseries]
      at (0,0.58) {Device state};
    \node[s3address, mapbox, minimum width=1.20cm] (map-reader-va)
      at (0,0.30) {Model VA};
    \node[s3model, mapbox, minimum width=1.68cm] (map-reader-gms)
      at (0,-0.70) {GMS pages \sThreeLock};
    \draw[s3mapping] (map-reader-va.south) --
      node[maplabel, right] {RO} (map-reader-gms.north);

    \node[maptitle] at (2.85,1.42) {(c) Prepared for Snapshot};
    \draw[s3panel] (1.90,-0.05) rectangle (3.80,1.15);
    \node[font=\tiny\bfseries] at (2.85,0.95) {Host process};
    \draw[s3devicegroup] (2.07,0.06) rectangle (3.63,0.72);
    \node[font=\fontsize{5.1}{5.5}\selectfont\bfseries]
      at (2.85,0.58) {Device state};
    \node[s3address, mapbox, minimum width=1.20cm]
      at (2.85,0.30) {Model VA};
    \node[s3model, mapbox, minimum width=1.68cm]
      at (2.85,-0.70) {GMS pages \sThreeLock};

    \draw[s3flow] (-1.84,0.84) -- node[maplabel, above] {Commit} (-1.01,0.84);
    \draw[s3flow] (1.01,0.86) -- node[maplabel, above] {Detach} (1.84,0.86);
    \draw[s3flow] (1.84,0.48) -- node[maplabel, below] {Re-attach} (1.01,0.48);
  \end{tikzpicture}
  \caption{GMS mappings with a Snapshot. A loader populates the
  GMS pages through an exclusive read-write mapping. Commit makes the pages
  immutable, and the engine reconnects with a read-only mapping.
  To prepare the engine for checkpoint, the GMS is detached from it, where the mapping and its imported handles are destroyed, but the virtual addresses are maintained.
  After restore, the process is reversed and the GMS is re-attached.
}
  \label{fig:gms-publication-reattachment}
\end{figure}
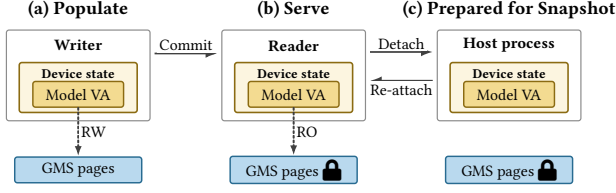

Composing Snapshots with GMS requires preserving model virtual
addresses embedded in runtime state while excluding the underlying model
allocations from the checkpoint. Before capture, the engine synchronizes,
unmaps each GMS range, releases its imported handle, and retains the
virtual-address reservation (Figure~\ref{fig:gms-publication-reattachment}).
GMS retains ownership of the physical allocations, so committed model state
remains resident and other readers are unaffected.

The resulting checkpoint contains the virtual addresses of the model tensors but not their underlying physical allocations, mappings, or contents~\cite{cuda-checkpoint}. After a
device-preserving failure, the restored engine imports the surviving GMS
allocations and maps them read-only at the retained addresses, avoiding model
restoration entirely. After a device-invalidating failure, the model state is
reconstructed in fresh GMS allocations in parallel with runtime restoration,
and the two are reattached once both complete. In either case, the restored
runtime requires compatible model layout, contents, permissions, and virtual
addresses~\cite{nvidia-cuda-graphs,nvidia-cuda-vmm}.

The two mechanisms remove complementary recovery work. Snapshots
replace engine initialization with runtime restoration, while GMS excludes
model state from the checkpoint and eliminates model restoration while its
resident state survives. Even when GMS state does not survive, repopulating
it \emph{in parallel} with runtime restoration shortens recovery relative to a
Snapshot alone, because current runtime-state restoration does not
saturate storage bandwidth. Snapshot restoration nevertheless remains
on the recovery critical path.

\subsection{Shadow Engines}
\label{sec:design-reattach}

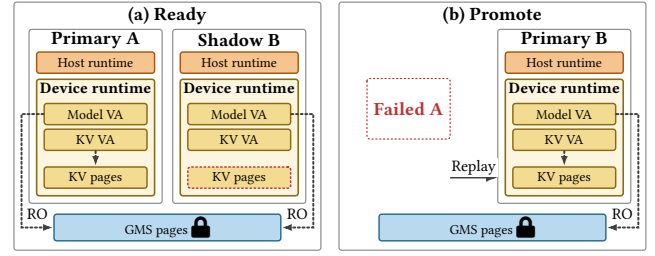
\begin{figure}[t]
  \centering
  \begin{tikzpicture}[
    stbox/.style={minimum width=1.36cm, minimum height=0.28cm,
      inner sep=1pt, font=\fontsize{5.2}{5.6}\selectfont},
    stwide/.style={minimum width=1.56cm, minimum height=0.28cm,
      inner sep=1pt, font=\fontsize{5.2}{5.6}\selectfont},
    sttitle/.style={font=\scriptsize\bfseries},
    stmaplabel/.style={font=\tiny, fill=white, inner sep=0.7pt}
  ]
\newcommand{\stEngine}[3]{\draw[s3panel] (-0.88,0) rectangle (0.88,-2.30);
      \node[sttitle] at (0,-0.14) {#2};
      \node[s3host, stwide] at (0,-0.44) {Host runtime};
      \draw[s3devicegroup] (-0.78,-0.66) rectangle (0.78,-2.20);
      \node[font=\tiny\bfseries] at (0,-0.77) {Device runtime};
      \node[s3address, stbox] (#1-model) at (0,-1.12) {Model VA};
      \node[s3address, stbox] (#1-kva) at (0,-1.46) {KV VA};
      \node[#3, stbox] (#1-kvp) at (0,-1.96) {KV pages};
    }

\begin{scope}[shift={(-2.15,0)}]
      \draw[s3panel] (-2.03,0) rectangle (2.03,-3.28);
      \node[sttitle] at (0,-0.18) {(a) Ready};
      \begin{scope}[shift={(-0.95,-0.36)}]
        \stEngine{ready-active}{Primary A}{s3device}
        \draw[s3mapping] (ready-active-kva.south) -- (ready-active-kvp.north);
      \end{scope}
      \begin{scope}[shift={(0.95,-0.36)}]
        \stEngine{ready-shadow}{Shadow B}{s3devicemissing}
      \end{scope}
      \node[s3model, stwide, minimum width=3.00cm] (ready-gms)
        at (0,-2.98) {GMS pages \sThreeLock};
      \draw[s3mapping] (ready-active-model.west) -- (-1.93,-1.48) |-
        (ready-gms.west);
      \node[stmaplabel] at (-1.72,-2.80) {RO};
      \draw[s3mapping] (ready-shadow-model.east) -- (1.93,-1.48) |-
        (ready-gms.east);
      \node[stmaplabel] at (1.72,-2.80) {RO};
    \end{scope}

\begin{scope}[shift={(2.15,0)}]
      \draw[s3panel] (-2.03,0) rectangle (2.03,-3.28);
      \node[sttitle] at (0,-0.18) {(b) Promote};
      \draw[s3panel, s3absent] (-1.66,-1.01) rectangle (-0.56,-1.81);
      \node[sttitle, text=sThreeFailure] at (-1.11,-1.41) {Failed A};
      \begin{scope}[shift={(0.95,-0.36)}]
        \stEngine{failover}{Primary B}{s3device}
        \draw[s3mapping] (failover-kva.south) -- (failover-kvp.north);
      \end{scope}
      \draw[s3flow] (-0.56,-2.32) -- node[stmaplabel, above] {Replay}
        (0.07,-2.32);
      \node[s3model, stwide, minimum width=3.00cm] (failover-gms)
        at (0,-2.98) {GMS pages \sThreeLock};
      \draw[s3mapping] (failover-model.east) -- (1.93,-1.48) |-
        (failover-gms.east);
      \node[stmaplabel] at (1.72,-2.80) {RO};
    \end{scope}
  \end{tikzpicture}
  \caption{Shadow Engine recovery. Primary and shadow share read-only GMS model state; only the primary holds KV pages. After failure, the shadow installs fresh KV pages and serves replayed requests. }
\Description{Two side-by-side panels compare the same active and shadow
  engines before and after recovery. Initially, both model virtual addresses
  have dotted read-only mappings to the same GMS pages. The primary engine has
  KV pages, while the shadow retains only its KV virtual address. After the
  primary engine fails, the promoted shadow retains its runtime and read-only
  model mapping, adds KV pages, and admits replayed requests.}
  \label{fig:shadow-lifecycle}
\end{figure}

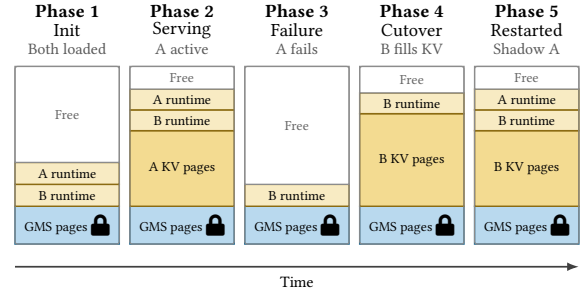
\begin{figure}[t]
  \centering
  \begin{tikzpicture}[
    memtitle/.style={font=\scriptsize\bfseries},
    memphase/.style={font=\scriptsize},
    memnote/.style={font=\tiny, text=black!65},
    memlabel/.style={font=\fontsize{5.2}{5.6}\selectfont, align=center},
    memfree/.style={draw=black!45, fill=white, line width=0.5pt},
    memruntime/.style={draw=sThreeDeviceEdge, fill=sThreeDevice!55,
      line width=0.55pt},
    memkv/.style={draw=sThreeDeviceEdge, fill=sThreeDevice,
      line width=0.55pt},
    memmodel/.style={draw=sThreeModelEdge, fill=sThreeModel,
      line width=0.55pt}
  ]
\foreach \x/\ttl/\phs/\nte in {
      -3.04/Phase 1/Init/Both loaded,
      -1.52/Phase 2/Serving/A active,
       0.00/Phase 3/Failure/A fails,
       1.52/Phase 4/Cutover/B fills KV,
       3.04/Phase 5/Restarted/Shadow A} {
      \node[memtitle] at (\x,3.09) {\ttl};
      \node[memphase] at (\x,2.83) {\phs};
      \node[memnote] at (\x,2.58) {\nte};
    }

    \begin{scope}[yscale=0.5]
\draw[memmodel] (-3.73,0.00) rectangle (-2.35,1.00);
    \node[memlabel] at (-3.04,0.50) {GMS pages \sThreeLock};
    \draw[memruntime] (-3.73,1.00) rectangle (-2.35,1.58);
    \node[memlabel] at (-3.04,1.29) {B runtime};
    \draw[memruntime] (-3.73,1.58) rectangle (-2.35,2.16);
    \node[memlabel] at (-3.04,1.87) {A runtime};
    \draw[memfree] (-3.73,2.16) rectangle (-2.35,4.70);
    \node[memlabel, text=black!60] at (-3.04,3.43) {Free};
    \draw[s3panel] (-3.73,0.00) rectangle (-2.35,4.70);

\draw[memmodel] (-2.21,0.00) rectangle (-0.83,1.00);
    \node[memlabel] at (-1.52,0.50) {GMS pages \sThreeLock};
    \draw[memkv] (-2.21,1.00) rectangle (-0.83,3.00);
    \node[memlabel] at (-1.52,2.00) {A KV pages};
    \draw[memruntime] (-2.21,3.00) rectangle (-0.83,3.55);
    \node[memlabel] at (-1.52,3.275) {B runtime};
    \draw[memruntime] (-2.21,3.55) rectangle (-0.83,4.10);
    \node[memlabel] at (-1.52,3.825) {A runtime};
    \draw[memfree] (-2.21,4.10) rectangle (-0.83,4.70);
    \node[memlabel, text=black!60] at (-1.52,4.40) {Free};
    \draw[s3panel] (-2.21,0.00) rectangle (-0.83,4.70);

\draw[memmodel] (-0.69,0.00) rectangle (0.69,1.00);
    \node[memlabel] at (0,0.50) {GMS pages \sThreeLock};
    \draw[memruntime] (-0.69,1.00) rectangle (0.69,1.58);
    \node[memlabel] at (0,1.29) {B runtime};
    \draw[memfree] (-0.69,1.58) rectangle (0.69,4.70);
    \node[memlabel, text=black!60] at (0,3.14) {Free};
    \draw[s3panel] (-0.69,0.00) rectangle (0.69,4.70);

\draw[memmodel] (0.83,0.00) rectangle (2.21,1.00);
    \node[memlabel] at (1.52,0.50) {GMS pages \sThreeLock};
    \draw[memkv] (0.83,1.00) rectangle (2.21,3.45);
    \node[memlabel] at (1.52,2.225) {B KV pages};
    \draw[memruntime] (0.83,3.45) rectangle (2.21,4.00);
    \node[memlabel] at (1.52,3.725) {B runtime};
    \draw[memfree] (0.83,4.00) rectangle (2.21,4.70);
    \node[memlabel, text=black!60] at (1.52,4.35) {Free};
    \draw[s3panel] (0.83,0.00) rectangle (2.21,4.70);

\draw[memmodel] (2.35,0.00) rectangle (3.73,1.00);
    \node[memlabel] at (3.04,0.50) {GMS pages \sThreeLock};
    \draw[memkv] (2.35,1.00) rectangle (3.73,3.00);
    \node[memlabel] at (3.04,2.00) {B KV pages};
    \draw[memruntime] (2.35,3.00) rectangle (3.73,3.55);
    \node[memlabel] at (3.04,3.275) {B runtime};
    \draw[memruntime] (2.35,3.55) rectangle (3.73,4.10);
    \node[memlabel] at (3.04,3.825) {A runtime};
    \draw[memfree] (2.35,4.10) rectangle (3.73,4.70);
    \node[memlabel, text=black!60] at (3.04,4.40) {Free};
    \draw[s3panel] (2.35,0.00) rectangle (3.73,4.70);
    \end{scope}

    \draw[s3flow] (-3.73,-0.30) -- node[below, font=\tiny] {Time} (3.73,-0.30);
  \end{tikzpicture}
  \caption{Device-memory occupancy during shadow recovery. After primary A fails, shadow B retains its runtime, installs KV pages, and serves while A restarts as the new shadow.}
  \Description{Five memory bars show initialization, serving, failure, cutover,
  and restart. Shared GMS pages persist throughout. Engine A's runtime and KV
  pages disappear at failure. Engine B retains its runtime, adds KV pages during
  cutover, and later runs beside the restarted Shadow Engine A.}
  \label{fig:shadow-memory-timeline}
\end{figure}

A Shadow Engine removes runtime restoration from the recovery path by keeping
a second initialized runtime on the same GPUs. Before serving begins, the
shadow initializes its host and device runtime, maps the committed model state
read-only through GMS, and then parks. Because it executes in a separate CUDA
context, its runtime remains intact after a device-preserving failure of the
primary engine.

The shadow preserves its CUDA context, communicators, fixed buffers, kernels,
CUDA Graphs, and the virtual-address layout they depend on. \textit{It requires neither
a second copy of the model nor physical KV pages: model pages are shared
through GMS, while the unused KV pool retains only its virtual-address
reservations.} Thus, the shadow's additional GPU footprint is limited to
engine-private runtime state.

Figure~\ref{fig:shadow-lifecycle} illustrates the resulting recovery path. The
primary engine alone holds KV pages and request-specific state, while both
engines retain read-only mappings of the same GMS model state. If the active
engine fails, the loss of its reader lease does not disturb the shadow's
mapping. The shadow can then install fresh KV pages at the retained addresses,
recreate the resources tied to those pages, and join the serving system once
its GPU state is complete. Because neither request records nor their KV values
survive the failure, the requests themselves must still be replayed.

After promotion, a new shadow is prepared while the promoted engine serves
(Figure~\ref{fig:shadow-memory-timeline}). We preferentially replenish it from a
Snapshot, avoiding exercising the much slower warm restart path.
Furthermore, Snapshots implicitly serialize CUDA graphs. Capturing them anew would require eager warmup execution that contends with the primary engine for GPU resources and decreases its serving throughput.
Once restored, the new shadow reattaches to the existing GMS model state and parks.

Replenishment also determines resilience to closely spaced failures. If another
failure occurs before a new shadow is ready, recovery falls back to GMS with
a warm restart or Snapshot restoration, placing runtime recovery back
on the critical path. Faster checkpoint-based replenishment therefore shortens
this vulnerability window. Additional shadows can eliminate it at the cost of
duplicating more runtime state.

\section{Implementation}
\label{sec:impl}

Our implementation consists of the node-local \snapshotrepo{} runtime, GMS,
and Shadow Engines, integrated with vLLM, SGLang, and TensorRT-LLM. The
current implementation replays interrupted requests rather than preserving
request state. This is not a design constraint: preserving request state is
compatible with \dynamo{}, but requires upstream changes to request scheduling
and KV-cache management and is left as ongoing future work (Section~\ref{sec:discussion}).
Our recovery implementation in \dynamo{} consists of 23k lines of Python, Go, and C.

\noindent\textbf{\textit{Snapshotting.}}\label{sec:impl-checkpoint}
A privileged node agent enters the engine container's namespaces and uses Checkpoint/Restore In Userspace (CRIU)
to checkpoint the host process tree and \texttt{cuda-checkpoint} for CUDA
state~\cite{criu,cuda-checkpoint,criugpu-arxiv25}. We contributed upstream CRIU
changes that parallelize host-memory page restoration speeding it up by 7.9$\times$ for a 129\,GiB checkpoint.

\texttt{cuda-checkpoint} preserves allocation contents but does not reconstruct
peer mappings or multicast objects required by multi-GPU engines. We interpose
on CUDA VMM and multicast APIs using an \texttt{LD\_PRELOAD} shim that records
these mappings. Before capture, the shim removes peer mappings, imported
handles, and multicast resources. After restore, it reconstructs them at their
original virtual addresses. The underlying handles may change, but preserving
virtual addresses keeps previously captured CUDA Graphs valid~\cite{nvidia-cuda-graphs}.

The engine and node agent coordinate through marker files in a pod-local
Kubernetes \texttt{emptyDir}. After initialization, the engine releases its KV
pages and signals \emph{ready-for-snapshot}. The agent captures the checkpoint
and, after recovery, signals \emph{restore-complete}. The engine then reinstalls
KV pages and registers with the serving control plane.

\noindent\textbf{\textit{GMS.}}\label{sec:impl-gms}
One GMS server per GPU creates physical allocations with
\texttt{cuMemCreate} and retains their creator handles. Clients receive
ephemeral POSIX descriptors over Unix-domain sockets, import them, and map the
allocations with the requested permissions~\cite{nvidia-cuda-vmm}. A
GMS-backed PyTorch CUDA memory pool wraps model construction and weight
loading, capturing model allocations without engine-specific enumeration.

To amortize VMM operations, the pool uses a first-fit suballocator over
2\,GiB slabs, with larger dedicated slabs for individual allocations exceeding
2\,GiB. Export/import, mapping, and permission setup therefore occur per slab
rather than per tensor, at the cost of internal fragmentation. Before commit,
the client records tensor layouts, removes loading scratch space, and moves
mutable state to private allocations. For checkpoint recovery with a cold
GMS, a separate read-write client repopulates model state from storage in
parallel with process restoration and commits it before the restored engine
reattaches.

\noindent\textbf{\textit{Shadow Engines.}}\label{sec:impl-shadow}
The operator launches engine candidates concurrently, whose leaders contend on
a shared kernel file lock. The winner becomes the primary engine. When it fails, the kernel
releases its descriptor and a shadow acquires the lock and becomes the new primary. The engine's existing
control plane coordinates follower ranks, extending the protocol across
multi-rank and multi-node replicas. Each shadow maps GMS weights read-only
and reserves its final KV virtual addresses. Upon promotion, it installs fresh
KV backing, activates its ranks, and registers with the serving control plane.

\section{Evaluation}
\label{sec:eval}

\subsection{Experimental setup}
\label{sec:eval-experimental-setup}

We evaluate Dynamo's recovery mechanisms on an eight-GPU NVIDIA DGX B200 node configured similarly
to the \dynamocluster{}. Although the production cluster contains $\sim$900 GPUs,
recovery operates independently within each replica;
its recovery latency and resource overhead are independent of the number of replicas in the serving fleet.
Our single-replica experiments therefore exercise the complete recovery path of a production-sized replica,
while Section~\ref{sec:eval-gpu-hours} separately evaluates fleet-level impact
by replaying the 18-week trace collected from the \dynamocluster{}.

We evaluate vLLM (v0.27.1) and SGLang (v0.5.18) across four eight-GPU model
configurations: Qwen3.8-27B (BF16, TP8), DeepSeek-V4-Flash (NVFP4, TEP8),
GLM-5.2 (NVFP4, TEP8), and DeepSeek-V4-Pro (NVFP4, TEP8). Each replica
occupies all eight GPUs, with prefix caching and chunked prefill enabled and
device-memory utilization set to 0.8. Prefill and decode are co-located.

We evaluate device-preserving failures, which account for most failures
in our production trace (Section~\ref{sec:failures}). We inject a failure by
sending \texttt{SIGKILL} to all processes in the engine process tree.
Snapshots and GMS artifacts are stored
on a VAST-backed network file system mounted with \texttt{nconnect=16}.

\subsection{Recovery Latency}
\label{sec:eval-recovery}

\begin{table*}[t]
  \small
  \centering
	\caption{Recovery latency (seconds) and speedup over warm restart for different models and frameworks.}
  \label{tab:recovery-spectrum}
  \setlength{\tabcolsep}{4pt}
  \begin{tabular*}{\textwidth}{@{\extracolsep{\fill}}lrrrr@{}}
    \toprule
    \textbf{Configuration} & \textbf{Qwen3.8-27B} & \textbf{DSV4-Flash} &
      \textbf{GLM-5.2} & \textbf{DSV4-Pro} \\
    \midrule
    \multicolumn{5}{@{}l}{\textbf{vLLM}} \\
    Warm restart & 106.4\,{\scriptsize(1.0$\times$)} & 109.5\,{\scriptsize(1.0$\times$)} & 144.4\,{\scriptsize(1.0$\times$)} & 156.5\,{\scriptsize(1.0$\times$)} \\
    GMS & 80.0\,{\scriptsize(1.3$\times$)} & 85.3\,{\scriptsize(1.3$\times$)} & 82.0\,{\scriptsize(1.8$\times$)} & 91.9\,{\scriptsize(1.7$\times$)} \\
    Snapshot & 40.5\,{\scriptsize(2.6$\times$)} & 60.0\,{\scriptsize(1.8$\times$)} & 131.7\,{\scriptsize(1.1$\times$)} & 189.7\,{\scriptsize(0.8$\times$)} \\
    Snapshot + cold GMS & 21.9\,{\scriptsize(4.9$\times$)} & 26.8\,{\scriptsize(4.1$\times$)} & 26.8\,{\scriptsize(5.4$\times$)} & 50.0\,{\scriptsize(3.1$\times$)} \\
    Snapshot + warm GMS & 16.6\,{\scriptsize(6.4$\times$)} & 24.1\,{\scriptsize(4.5$\times$)} & 24.9\,{\scriptsize(5.8$\times$)} & 25.7\,{\scriptsize(6.1$\times$)} \\
    \bfseries GMS + Shadow Engine & \bfseries 4.6\,{\scriptsize(23.1$\times$)} & \bfseries 5.6\,{\scriptsize(19.6$\times$)} & \bfseries 5.6\,{\scriptsize(25.8$\times$)} & \bfseries 5.4\,{\scriptsize(29.0$\times$)} \\
    \midrule
    \multicolumn{5}{@{}l}{\textbf{SGLang}} \\
    Warm restart & 79.6\,{\scriptsize(1.0$\times$)} & 95.7\,{\scriptsize(1.0$\times$)} & 144.6\,{\scriptsize(1.0$\times$)} & 151.2\,{\scriptsize(1.0$\times$)} \\
    Snapshot + cold GMS & 37.1\,{\scriptsize(2.1$\times$)} & 31.1\,{\scriptsize(3.1$\times$)} & 38.0\,{\scriptsize(3.8$\times$)} & 55.2\,{\scriptsize(2.7$\times$)} \\
    Snapshot + warm GMS & 33.3\,{\scriptsize(2.4$\times$)} & 31.1\,{\scriptsize(3.1$\times$)} & 35.3\,{\scriptsize(4.1$\times$)} & 29.8\,{\scriptsize(5.1$\times$)} \\
    \bfseries GMS + Shadow Engine & \bfseries 6.2\,{\scriptsize(12.8$\times$)} & \bfseries 6.2\,{\scriptsize(15.4$\times$)} & \bfseries 5.7\,{\scriptsize(25.4$\times$)} & \bfseries 5.3\,{\scriptsize(28.5$\times$)} \\
    \bottomrule
  \end{tabular*}
\end{table*}

Table~\ref{tab:recovery-spectrum} compares recovery latency across mechanisms implemented in \dynamo{}.
For Shadow Engines, we measure from failure injection until the promoted
engine registers with the discovery plane and is ready to serve. For all other
configurations, we measure from container start until the same point.

\begin{figure}[!b]
  \centering
  \includegraphics[width=\linewidth]{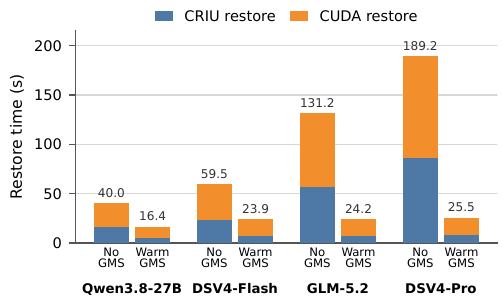}
  \caption{vLLM Snapshot restoration time without GMS and with a warm GMS, where model weights are already resident, excluding KV cache reallocation and discovery registration time.}
  \label{fig:checkpoint-restore-breakdown}
\end{figure}

We use a warm restart as the baseline, with model weights already in the host
page cache and compilation artifacts cached. This is optimistic relative to
production: most recoveries in the \dynamocluster{} take longer than even our slowest warm
restart (Figure~\ref{fig:gpu-hours}).

\dynamo{}'s GMS alone provides a 1.3--1.8$\times$ speedup by eliminating model loading,
but still executes the remaining engine initialization path. Snapshots
eliminate initialization but must restore captured device memory.
This yields a 2.6$\times$ speedup for Qwen3.8-27B, but scales poorly with model
size and is slower than the warm-restart baseline for the largest model we tested (DeepSeek-V4-Pro).
Figure~\ref{fig:checkpoint-restore-breakdown} shows that CUDA restoration
dominates this cost, as \texttt{cuda-checkpoint} copies device memory serially and
does not saturate PCIe bandwidth.

Combining Snapshots with a cold GMS loads weights into GMS \emph{in
parallel} with runtime-state restoration. This occurs on first startup or upon a device-invalidating failure.
The recovery latency is the slower of the two paths, yielding a
2.1--5.4$\times$ speedup over warm restart.
Snapshots with a warm GMS, where weights are \emph{already resident},
remove GMS population from the recovery path and yield a 2.4--6.4$\times$
speedup over warm restart. The remaining variation is dominated by restoring
the reduced Snapshot artifact and by second-restore page-cache warming, rather
than by loading model weights (Figure~\ref{fig:checkpoint-restore-breakdown}).
Shadow Engines remove runtime restoration from the
critical path as well, as recovery reduces to installing fresh KV backing and
registering the engine for serving. Consequently, recovery completes in
4.6--6.2\,s across all models and frameworks, a 12.8--29.0$\times$ speedup over
warm restart.

\subsection{Device-Memory Cost of a Shadow Engine}
\label{sec:eval-memory}

\begin{figure}[t]
  \centering
  \includegraphics[width=\linewidth]{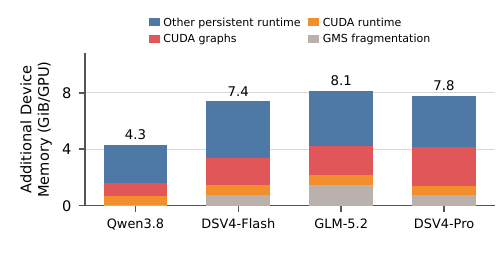}
  \caption{Breakdown of the additional steady-state device memory occupied by one vLLM
  Shadow Engine.}
  \label{fig:shadow-memory}
\end{figure}

Shadow Engines trade device memory for recovery latency. Because the shadow
shares model state through GMS and holds no physical KV pages while parked,
its additional footprint consists primarily of engine-private runtime state,
reducing the memory available to the active engine's KV cache.

Figure~\ref{fig:shadow-memory} decomposes this overhead into GMS
fragmentation, CUDA runtime, CUDA Graphs, and other persistent runtime state.
We measure total overhead as the difference in \texttt{nvidia-smi} device
memory between deployments with and without a shadow. GMS fragmentation corresponds to the slab allocations' unused portions that is unrelated to the shadow; CUDA runtime contributes
718 MiB, measured using a minimal CUDA process (empty CUDA program)
that initializes the runtime; and vLLM reports CUDA Graph memory during capture.
The remainder includes persistent communication and execution buffers such as NCCL and FlashInfer.

The shadow adds 4.3 GiB/GPU for Qwen3.8-27B and 7.4--8.1 GiB/GPU for the
three larger models. In contrast, the shared model state grows from
8 to 124 GiB/GPU. Thus, shadow overhead is largely independent of model
size, since model weights and KV-cache backing are not duplicated.

\subsection{Interference During Shadow Replenishment}
\label{sec:eval-replenishment}

\begin{figure}[t]
  \centering
  \includegraphics[width=\linewidth]{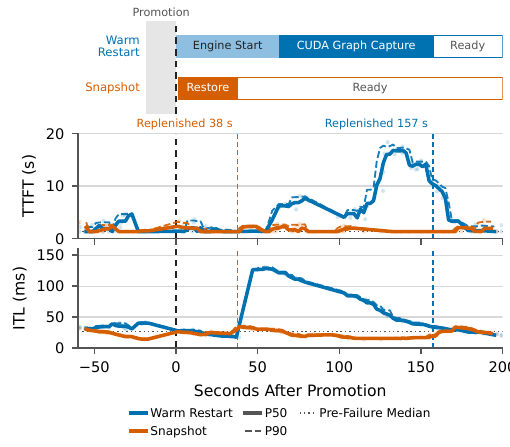}
   \caption{Shadow-replenishment interference during serving (vLLM GLM-5.2).
   Curves show 10-s sliding p50/p90 latencies. Dashed lines mark replenishment completion.
   The top rows show each replacement shadow's lifecycle.}
  \Description{A two-row replacement lifecycle sits above two latency panels.
  A warm restart takes 157 seconds and overlaps a period of elevated request
  latency; Snapshot restoration finishes in 38 seconds and the
  observed request latencies remain near their pre-failure levels.}
  \label{fig:replenishment}
\end{figure}

We evaluate whether Snapshots replenish a consumed shadow faster
and with less serving interference than a warm restart
(Section~\ref{sec:design-reattach}). We run vLLM with GLM-5.2 and TEP8 parallelism, promote its
shadow after failure, and replenish a new shadow on the same GPUs while the
promoted engine serves 32K/1K input/output requests. Requests arrive as a
Poisson process at 0.35 requests/s, a rate that a healthy replica sustains
without queuing.

Figure~\ref{fig:replenishment} shows that Snapshot restoration
replenishes the shadow in 38s, versus 157s for a warm restart. Of the
157s, 94s is spent capturing CUDA Graphs. This initialization competes with
the serving engine for GPU execution: prefill throughput drops from 9,600 to
6,400 tokens/s, while median TTFT and ITL increase from 1.33s and 25ms
before failure to 5.96s and 43ms for requests issued during replenishment.
Latency recovers only after initialization completes.

On the other hand, Snapshot restoration causes no measurable serving interference:
TTFT and ITL remain at their pre-failure values, and the waiting queue never
exceeds one request. It also reduces the interval without a shadow from
157s to 38s, shortening the window in which a second failure must fall
back to a slower recovery path (Table~\ref{tab:recovery-spectrum}).

\subsection{Serving Impact of Losing a Replica}
\label{sec:eval-replica-loss}

We next evaluate how recovery latency affects the service when traffic from a
failed replica is redirected to the remaining replicas until recovery completes.
If this displaced traffic exceeds their spare capacity, the survivors become
overloaded. This can occur when deployments operate with limited headroom,
failures coincide with peak load, or multiple failures occur within a short recovery window.

We deploy two vLLM GLM-5.2 replicas with the workload from
Section~\ref{sec:eval-replenishment}, but double the arrival rate to
0.7 requests/s---a load sustainable by two replicas but not one. Requests are
distributed round-robin with distinct prefixes. After 450s of steady-state
serving, we terminate one replica.

We compare a warm restart against Shadow Engine recovery. The warm restart
uses cached weights, compiled kernels, and autotuning results, yet requires
148s to rejoin the service. The Shadow Engine resumes serving in 5.8s.
Both runs have identical pre-failure throughput, TTFT, and ITL.

\begin{figure}[t]
  \centering
  \includegraphics[width=\linewidth]{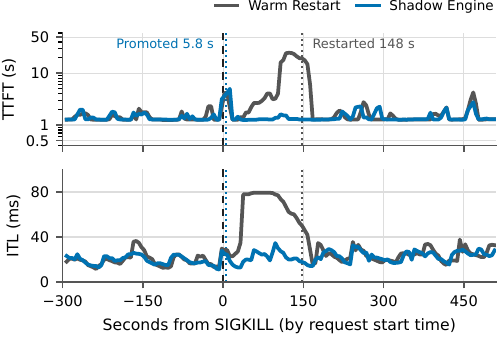}
  \caption{15-second rolling median time-to-first-token and inter-token latency
  during replica loss in a loaded two-replica vLLM GLM-5.2 deployment. Dotted lines mark
  recovery completion.}
  \Description{Two time-series panels compare warm restart and shadow recovery.
  Both curves rise to a median time to first token near 4 seconds for
  requests started in the 15 seconds after the failure. The shadow curves
  then return to their pre-failure values. The warm restart curve for time to
  first token climbs to about 4 seconds by 100 seconds, jumps to between 17
  and 25 seconds for requests started 110 to 160 seconds after the failure,
  and returns to 1.3 seconds at 165 seconds. Its inter-token latency rises to
  80 milliseconds for requests started 35 to 105 seconds after the failure
  and declines to about 40 milliseconds before the replica returns at 148
  seconds.}
  \label{fig:replica-failure}
\end{figure}

Figure~\ref{fig:replica-failure} shows the resulting serving impact. With
Shadow Engine recovery, only requests immediately following the failure see a
brief TTFT increase, and recovery completes before the surviving replica becomes
overloaded. With a warm restart, the survivor carries the full load
for 148s: median TTFT increases from 1.3s to approximately 20s, median
ITL from 23 to 80ms, and six requests fail, compared with one under shadow
recovery.

The dominant effect is admission pressure. With 32K input tokens and 1K output tokens per request,
each request occupies 33K KV tokens, limiting the survivor's 1.8M-token KV pool
to 55 concurrent sequences.
During the warm restart, this pool fills approximately 110s after failure,
forcing subsequent requests to wait for admission. TTFT therefore continues
to grow with the recovery window and remains elevated while the queue drains
after recovery. ITL is comparatively bounded because admitted requests decode
within the fixed concurrency limit. In contrast, the 5.8s shadow promotion
restores capacity well before KV exhaustion, despite reducing the primary engine's KV capacity by 9\% while parked (Section~\ref{sec:eval-memory}).

\subsection{Production GPU-Hour Savings}
\label{sec:eval-gpu-hours}
Our previous experiments show that Shadow Engines reduce recovery to under
7s. We next ask how much production GPU capacity this reduction would have
reclaimed in the \dynamocluster{} over the 18-week trace (Section~\ref{sec:failures}).

We perform a counterfactual analysis over the 4,443 device-preserving
failures identified in Section~\ref{sec:failures}, excluding the 1,319 ambiguous restarts. For each failure, we
retain its observed replica size and recovery duration, but replace the
recovery duration with 7s, conservatively rounding up the slowest Shadow
Engine promotion in Table~\ref{tab:recovery-spectrum}. The reclaimed capacity
for a failure is therefore the reduction in recovery time multiplied by the
number of GPUs in that replica. Summing across failures directly translates
the measured recovery improvement into production GPU-hours.

We evaluate only Shadow Engines because their measured promotion latency is
nearly independent of model size and framework. In contrast, GMS and Snapshot
recovery depend on the amount of model and runtime state and are
measured only for the configurations in Section~\ref{sec:eval-recovery}.
The analysis also excludes failure-detection time and any savings from reducing
overprovisioned capacity, and assumes that a shadow is available at each
failure and replenished off the critical path (Section~\ref{sec:eval-replenishment}).

\begin{figure}[t]
  \centering
  \includegraphics[width=\linewidth]{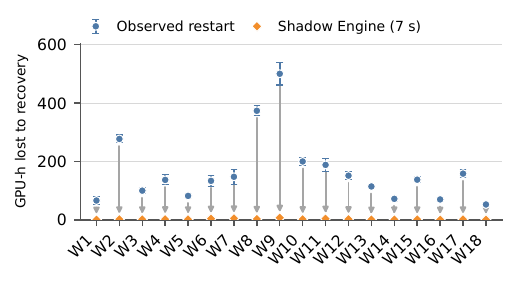}
  \caption{Weekly GPU-hours lost to recovery in the \dynamocluster{}, observed versus 7-s shadow promotion
  for device-preserving failures. Arrows show savings. Whiskers show timing uncertainty
  (Appendix~\ref{app:trace-timing}).}
  \Description{Chart over 18 weeks on a linear axis. Observed values sit
  between 50 and 500 GPU-hours per week, with an arrow from each pointing down
  to the shadow value between 0.8 and 7.5, nearly at zero.}
  \label{fig:gpu-hours-shadow}
\end{figure}

Figure~\ref{fig:gpu-hours-shadow} shows the result.
\textit{Shadow Engines alone would reclaim 2,339 GPU-hours lost across the 18-week production trace.
This corresponds to 98\% of the 2,375 GPU-hours lost to device-preserving recoveries, and 79\% of the 2,967 GPU-hours lost to all in-place recoveries, including ambiguous ones.}

\section{Limitations and Extensions}
\label{sec:discussion}

\paragraph{Host-resident shadow.}
A Shadow Engine consumes 4--8\,GiB/GPU (Section~\ref{sec:eval-memory}),
less than 5\% of device memory on NVIDIA B200 and equivalent to approximately 128K
tokens of KV capacity for a 64\,KiB/token model. We consider this a modest
cost for reducing recovery from minutes to under 7\,s, particularly when
serving is not KV-capacity bound. Moreover, Section~\ref{sec:eval-replica-loss}
shows that this reduction in KV capacity can be preferable to losing an entire
replica for minutes: despite a slightly smaller KV pool, the shadow restores
capacity before the surviving replica becomes saturated.

For deployments where KV capacity is the primary bottleneck, even this
overhead may be undesirable. An intermediate design could retain only the
shadow's host runtime state and checkpoint its device state, restoring the
latter with \texttt{cuda-checkpoint} on failover. This trades lower steady-state
memory for higher recovery latency, occupying a middle point between Snapshots
and Shadow Engines. We leave this design, together with optimizing
device-state restoration toward its bandwidth limit, to future work.

\paragraph{Large deployments.}
Our production trace and evaluation cover single-node replicas. Rack-scale
replicas with wide expert parallelism~\cite{deepseek-inference-overview,nvidia-gptoss-nvl72}
span many GPUs across multiple nodes, increasing the capacity lost when any
rank fails. Dynamo's recovery implementation already supports multi-node recovery: its shadow
election coordinates ranks across nodes, while GMS preserves model state
independently on each GPU (Section~\ref{sec:impl-shadow}). However, our
production failure characterization may not generalize to this setting.
Larger replicas introduce additional failure domains, including inter-node
fabric and hardware faults, which may increase the fraction of
device-invalidating failures. Quantifying this distribution requires
production traces from rack-scale deployments and remains future work.

\paragraph{Preserving KV state.}
The design accelerates recovery of serving capacity but currently reconstructs
request state after failure. Near-instantaneous resumption of in-flight requests
would additionally require preserving scheduler metadata, KV mappings, and KV
contents. Existing KV-offloading systems such as LMCache and kvcached~\cite{lmcache-arxiv25, kvcached1, kvcached2}
can preserve committed KV blocks in host memory or storage, allowing them to be
restored even after device-invalidating failures when the host remains
available. For device-preserving failures, GMS could instead own physical KV
pages, allowing KV contents to remain resident and be reattached after recovery.
These mechanisms provide complementary tiers of request-state resilience, but
require inference engines~\cite{vllm-sosp23,sglang-neurips24} to expose and
restore KV-cache and scheduler metadata. We leave this integration to future
work.

\paragraph{Fast engine reconfiguration.}
The design can also accelerate engine reconfiguration. Changes to scheduling
policies, CUDA graph shapes, or other runtime configuration currently require
reinitializing the engine~\cite{vllm-sosp23,sglang-neurips24}. When the
reconfiguration preserves the rank-local weight shards, GMS allows a new
engine configuration to reuse the active engine's committed model state.
A Shadow Engine can therefore be initialized with the new configuration and
promoted without reloading model weights. Snapshots can similarly
capture frequently used configurations for later restoration. Thus, the same
separation of model and runtime state that accelerates failure recovery can
also reduce reconfiguration latency.

\section{Related Work}
\label{sec:related}

\paragraph{Fault-tolerant LLM serving.}
Training systems bound lost progress through checkpointing, redundancy, or
pipeline adaptation~\cite{checkfreq-fast21,gemini-sosp23,bytecheckpoint-arxiv24,bamboo-nsdi23,oobleck-sosp23,recycle-sosp24,pccheck-asplos25}.
D\'ej\`aVu, GhostServe, LUMEN, and Concordia preserve request progress through KV state~\cite{dejavu-icml24,ghostserve-mlsys26,lumen-arxiv26,concordia-arxiv26};
FailSafe, Tarragon, KevlarFlow, and ReviveMoE reconfigure parallelism around failed workers~\cite{failsafe-arxiv25,tarragon-arxiv26,kevlarflow-arxiv26,revivemoe-arxiv26};
SpotServe adapts to GPU preemption~\cite{spotserve-asplos24}.
These mechanisms complement Dynamo's recovery mechanisms, which target the loss of initialized
serving capacity (Section~\ref{sec:motivation}). Extending this recovery design to training is future work.

\paragraph{Classical recovery principles.}
Crash-only design, microreboot, and CuriOS separate recoverable state from
restartable components~\cite{crashonly-hotos03,microreboot-osdi04,curios-osdi08}.
Rio and Otherworld retain memory across OS crashes~\cite{rio-asplos96,otherworld-eurosys10};
shadow drivers conceal driver failures~\cite{shadowdrivers-osdi04}, and Remus
provides warm shadows~\cite{remus-nsdi08}.
Dynamo's recovery design applies state separation and shadow recovery to inference engines and GPU state.

\paragraph{External ownership of inference state.}
AnchorTP externalizes weights and KV caches for elastic parallelism; ServerlessLLM exposes loaded weights for cold starts~\cite{anchortp-date26,serverlessllm-osdi24};
KV offloading layers retain blocks across requests~\cite{lmcache-arxiv25,cachegen-sigcomm24,cachedattention-atc24}.
Torpor shares GPU runtimes through external executors~\cite{torpor-atc25}.
vAttention and GMLake separate virtual mappings from physical allocations~\cite{vattention-asplos25,gmlake-asplos24};
GPM provides GPU access to host persistent memory~\cite{gpm-asplos22}.
\dynamo{}'s GMS instead decouples volatile allocation ownership from engine lifetime;
\dynamo{} also restores runtime readiness.

\paragraph{GPU checkpoint and restore.}
CRIU, CUDA checkpointing, CRIUgpu, and Singularity restore process or GPU
state~\cite{criu,cuda-checkpoint,criugpu-arxiv25,singularity-arxiv22}.
HeteroCheckpoint coordinates host/device checkpoints~\cite{heterocheckpoint-dsn14};
CRUM uses a proxy process and CRAC a split-process architecture to checkpoint CUDA unified memory~\cite{crum-cluster18,crac-sc20}.
PhoenixOS, gCROP, and GCR reduce checkpoint/restore overhead~\cite{phoenixos-sosp25,gcrop-socc24,gcr-fast26}.
Dynamo's recovery design builds on these foundations for serving recovery and composition with GMS.

\paragraph{Fast startup and runtime materialization.}
VM restore and REAP prioritize working sets~\cite{fastrestore-vee11,reap-asplos21}.
PipeSwitch, BlitzScale, and HydraServe accelerate weight movement~\cite{pipeswitch-osdi20,blitzscale-osdi25,hydraserve-nsdi26};
Medusa, TIDAL, and Foundry restore CUDA graphs and GPU function state from runtime snapshots~\cite{medusa-asplos25,tidal-arxiv25,foundry-arxiv26}.
Dynamo's recovery design distinguishes device-preserving from device-invalidating failures to
reuse surviving allocations and initialized shadows. Device memory survival does not imply durability across resets. Accelerated reconstruction
remains complementary when state is lost.

\section{Conclusion}
\label{sec:conclusion}
LLM inference recovery need not reconstruct an engine from scratch. Our
production study shows that most failures are device-preserving: the
engine process fails, but the GPUs and reusable state on them survive.
Dynamo's recovery architecture exploits this asymmetry by decoupling initialized model and runtime
state from the lifetime of the engine process. Snapshots preserve
initialized runtimes, GMS preserves GPU-resident model state, and Shadow
Engines preserve a ready runtime on the same GPUs. Together, these mechanisms
progressively remove initialization from the recovery critical path while
retaining a durable fallback for device-invalidating failures.

Across multiple models and serving frameworks, \dynamo{} reduces recovery
from minutes to under 7\,s with 4--8\,GiB/GPU of shadow memory, prevents
prolonged overload when a replica fails, and would reclaim 2,339 of the 2,375
GPU-hours lost to device-preserving failure recovery in our 18-week trace from the \dynamocluster{}.
The implementation is open source as part of \dynamo{} and deployed in production, demonstrating that these mechanisms
integrate with real inference serving stacks agnostic to inference engines
like vLLM, SGLang, or TensorRT-LLM. More broadly, \textit{our results show
that fast inference resiliency is fundamentally a state-lifetime problem:
when initialized state survives independently of the process that created it,
restoring serving capacity becomes promotion and reattachment rather than
reinitialization.
}

\bibliographystyle{ACM-Reference-Format}

\clearpage

\appendix

\section{Failure Signals and Recovery Actions}
\label{app:failure-signals}

Table~\ref{tab:relevant-xids} lists the GPU-side fault signals relevant to
the recovery policy and the resident-state reuse decision each
implies~\cite{nvidia-xid-catalog,nvidia-gpu-error-containment,
nvidia-gpu-recovery-actions,nvidia-cuda-vmm}. The trace records XIDs as a
last-value gauge per GPU, which shows that a fault occurred but cannot
attribute it to a particular restart. We therefore exclude every restart on
a host that reported a reset-class or memory-integrity code in the same week.

\begin{table*}[t]
  \small
  \centering
  \caption{XID signals, resident-state reuse decision, and recovery action.}
  \label{tab:relevant-xids}
  \begin{tabular}{@{}p{0.06\textwidth}p{0.42\textwidth}p{0.12\textwidth}p{0.30\textwidth}@{}}
    \toprule
    \textbf{XID(s)} & \textbf{Meaning} & \textbf{Resident state} &
    \textbf{Recovery action} \\
    \midrule
    8 & GPU stopped processing work for the
    application (channel idle timeout). & Survives & Restart the replica. \\
    \addlinespace
    13 & Graphics-engine exception, typically an application fault such as an
    out-of-bounds access. & Survives & Restart the replica. \\
    \addlinespace
    31 & GPU MMU page fault, often from an illegal address access. & Survives
    & Restart the replica. \\
    \addlinespace
    43 & The application channel terminated after a software-induced fault
    and the GPU remains healthy. A preceding XID may identify the underlying fault.
    & Survives & Restart the replica. \\
    \addlinespace
    45 & The driver removed a channel with work still pending, typically
    after a preceding error or when its process was killed mid-kernel.
    & Survives & Restart the replica. \\
    \addlinespace
    11, 32, 69, 109, 145 & PBDMA, context-switch timeout, and related
    graphics-engine faults that leave the GPU unresponsive. & Lost & Reset the
    GPU. \\
    \addlinespace
    48 & Double-bit ECC error. & Lost & Reset the GPU. \\
    \addlinespace
    79 & The GPU is inaccessible over PCIe (``fallen off the bus''). & Lost &
    Reboot/repair the node. \\
    \addlinespace
    94 & Contained uncorrectable memory error, isolated to one application.
    & Lost & Reset the GPU. \\
    \addlinespace
    95 & Uncontained memory error affecting multiple applications. & Lost &
    Reset the GPU. \\
    \addlinespace
    119/120 & GSP RPC timeout or GSP error. & Lost & Reset the GPU and/or reboot
    the node. \\
    \bottomrule
  \end{tabular}
\end{table*}

\section{Timing Recoveries in the Trace}
\label{app:trace-timing}

The recovery durations in Figures~\ref{fig:gpu-hours}
and~\ref{fig:gpu-hours-shadow} come from the cluster's monitoring system,
which scrapes the Kubernetes restart counter and readiness condition of
every worker container every 15 seconds. We count only in-place container
restarts, which increment this counter. Worker pods run with
\texttt{restartPolicy: Always}, so a crashed engine restarts inside the same
pod, and the same pod, node, and GPUs are on both sides of the interval.
Pod replacement, which results from rollouts, configuration changes,
autoscaling, or eviction rather than from failures, does not increment the
counter and is not counted. A recovery starts at the first sample where the
restart counter has increased and ends at the first later sample where the
container is ready again. Its lost GPU-hours are that duration times the
replica's GPU limit. A first pass over 12-hour windows returns 2-minute samples and
enumerates the recoveries, counting a counter jump of several restarts
inside one sample as one idle interval. A second pass re-queries every
2-hour window that contains a recovery at 30-second resolution and re-times
each one, keeping the 2-minute timing for the 12\% whose readiness series
has gaps or contradicts the coarse reading. Pods that served no requests that week are excluded, as are pods with 24
or more restarts in a week. The restart distribution is bimodal: serving
replicas restart a few times per week, whereas a small number of
misconfigured deployments crash-loop during initialization without ever
serving, some restarting more than a hundred times in a day. The threshold
sits between these two populations. Excluding the loops is also
conservative, since an engine that never finishes initialization loses no
serving capacity that a shadow could restore. The three
device-invalidating episodes did not restart in place and therefore do not
appear.
This enumerates 5{,}762 recoveries, of which 1{,}319 (592 GPU-hours) are
excluded as ambiguous: 810 ran on hosts whose GPUs reported a reset-class or
memory-integrity XID that week, 506 cannot be attributed to a node by the
monitoring data, 38 saw the node's allocatable GPU count drop during the
recovery, and 2 were part of a node-wide restart burst. No recovery followed
a node reboot. The 98 remaining recoveries that ended in another crash
before the container was ready, about 6\% of the lost GPU-hours, are
included with the next restart as their end.

We cross-checked the timing against Kubernetes events, which are retained
for the whole trace with timestamps to the second, on a sample of 108
recoveries, six per week, from pods that restarted at most three times in
the trace. The kubelet's container-start event precedes our start by a
median of 15 seconds and by under 32 seconds in 90\% of cases. Kubernetes
records probe failures but not the success that ends them, so an
event-based end can only be bracketed to one probe period, and where the
failure stream is complete the bracket agrees with our end within 30
seconds. The shadow latency of Section~\ref{sec:eval-recovery} ends at
registration rather than at the probe. The 7-second assumption in
Section~\ref{sec:eval-gpu-hours} sits above the slowest measured promotion
plus its 1-second probe period, so the two ends are comparable. Two biases run toward late
ends. The observed end trails registration by up
to one probe period of 10 seconds, and about 1.5\% of readiness series have
gaps that the monitoring system fills by interpolation. Together they can
overstate the trace total by at most 77 GPU-hours, or 3\%. Random timing
error averages out over 4{,}443 recoveries to about 4 GPU-hours. Timing
every recovery at the short end of its sampling window, and no shorter than
one 15-second scrape, gives a lower bound of 2{,}174 GPU-hours lost, of
which the replay of Section~\ref{sec:eval-gpu-hours} reclaims 2{,}138, again
98\%.

\end{document}